\documentclass[twocolumn,pre]{revtex4}
\usepackage{amsmath,amssymb}
\usepackage{graphicx,pstricks}
\usepackage{url}
\usepackage{bm}
\include{epsf}

\def\(({\left(}
\def\)){\right)}                       
\def\[[{\left[}
\def\]]{\right]}

\renewcommand{\>}{\rangle}
\newcommand{\beq}{\begin{equation}}
\newcommand{\eeq}{\end{equation}}
\newcommand{\bea}{\begin{eqnarray}}
\newcommand{\eea}{\end{eqnarray}}

\begin{document}
\title{R\'enyi entropy, abundance distribution and the equivalence of ensembles}

\author{Thierry Mora$^{1}$, Aleksandra M. Walczak$^{2}$
}

\affiliation{$^{1}$ Laboratoire de physique statistique,
    CNRS, UPMC and \'Ecole normale sup\'erieure, 24, rue Lhomond,
    Paris, France}
\affiliation{$^{2}$ Laboratoire de physique th\'eorique,
    CNRS, UPMC and \'Ecole normale sup\'erieure, 24, rue Lhomond,
    Paris, France}

\begin{abstract}
 % abstract
Distributions of abundances or frequencies play an important role in many fields of science, from biology to sociology, as does the R\'enyi entropy, which measures the diversity of a statistical ensemble. We derive a mathematical relation between the abundance distribution and the R\'enyi entropy, by analogy with the equivalence of ensembles in thermodynamics. The abundance distribution is mapped onto the density of states, and the R\'enyi entropy to the free energy. The two quantities are related in the thermodynamic limit by a Legendre transform, by virtue of the equivalence between the micro-canonical and canonical ensembles. In this limit, we show how the R\'enyi entropy can be constructed geometrically from rank-frequency plots. This mapping predicts that non-concave regions of the rank-frequency curve should result in kinks in the R\'enyi entropy as a function of its order. We illustrate our results on simple examples, and emphasize the limitations of the equivalence of ensembles when a thermodynamic limit is not well defined. Our results help choose reliable diversity measures based on the experimental accuracy of the abundance distributions in particular frequency ranges.

\end{abstract}

\maketitle

 % maintext
As an increasing number of large datasets are becoming available in a variety
of fields, one often turns to reduced statistics that can capture
important properties of the system, or help detect deviations from our
expectations. Distributions of abundances have
proven useful as such statistics, and have been used in many different contexts, from biology to linguistics, astrophysics and sociology. This notion is best explained when counting biological species from a sample. Say that species 1 was observed $n_1$ times, species 2 $n_2$ times, etc. The abundance distribution discards information about the {\em identity} of the sampled species, and focuses on the distribution of the counts themselves $n_1$, $n_2$, etc. This notion is very general and extends well beyond ecology. Counts can refer to the number of times a word is used in a text, to the number of people living in a given city, to the occurence of specific spiking patterns in a population of neurons, or to the abundance of specific lymphocyte clones in the immune system, to give but a few examples.

An equivalent way of representing the abundance distribution is to
order the counts from largest to smallest, and plot them as a function
of their rank in this ordering. For example, in the English language,
one can order words by their frequency of occurence, and study how
this frequency decreases with the rank. In 1949 Zipf observed that
this dependency roughly followed a power law \cite{Zipf}, and similar
observations have later been made in a variety of contexts
\cite{Newman2005}. Because of the ubiquity of these power laws \cite{Schwab2014}, frequency-versus-rank plots are commonly represented on a double logarithmic scale.

Abundance distributions can contain useful, albeit indirect,
information about the underlying process at work in the system. In
ecology, they are used as a diagnostic tool for detecting deviations
from the prediction of a neutral model of population dynamics \cite{Volkov2003,Chisholm2010}. The Yule speciation process
\cite{Yule,Simon}, called the preferential attachment process in the
context of networks \cite{Barabasi1999}, also predicts a specific form
for the abundance distribution, which is consistent with Zipf's law in
some limit \cite{Simon}. The abundance distribution of spike patterns
in the retina has been used to study the critical properties of the
underlying neural network \cite{Tkacik2015}, and a similar analysis was
performed on small patches of natural images \cite{Stephens2013}.
The distribution of sizes of lymphocyte clones in the immune system
also seems to generically follow power-laws, which puts constraints on
the rules of their population dynamics \cite{Desponds2016}.

Abundance distributions are closely related to the notion of
diversity. Diversity can be defined in a number of ways: total number
of types in the distribution, Shannon's entropy \cite{Shannon}, Simpson's diversity
index, {\em etc}. It has long been realized \cite{Hill1973} that these
different kinds of diversity can all be brought under the common
definition of the R\'enyi entropy \cite{Renyi1961}. This quantity,
which depends on a single parameter called order, generalizes 
Shannon's and Gibbs' entropy. It
is commonly used in ecology to quantify diversity, but has also received
increasing attention in
condensed matter in the context of quantum entanglement \cite{Hastings2010}.

Here we show how the R\'enyi entropy can be geometrically constructed
from the abundance distribution in the thermodynamic limit. This
construction allows one to visualize graphically how different
measures of diversity arise from a given abundance distribution. It
indicates which are the abundances that are determinant in each
diversity measure, and gives a visual assessment of when to trust the
estimate of these measures.

Our result relies on the framework of statistical mechanics, piecing
together previous observations. The equivalence between rank-frequency curves and the micro-canonical entropy has been previously reported in \cite{Mora2011a}. The link between micro-canonical entropy and free energy is a classical result of statistical mechanics, known as the equivalence of ensembles \cite{Touchette2004}. The mapping between free energy and the R\'enyi entropy has been pointed out recently \cite{Baez2011}.
By bringing these results in a common framework, we hope to clarify
the correspondance between abundance distributions, the density of states and R\'enyi entropies, and propose a straightforward geometric method for assessing diversity directly from the abundance distribution represented in an appropriate manner.

\section*{R\'enyi entropy and free energy}

Let us define a probability distribution $p(s)$, where $s$ is a state
or a type that can take a
discrete number of values. $p(s)$ is a relative abundance, or a
frequency, so that $\sum_s p(s)=1$. The variable $s$ can be a spin
configuation of a large system, a species, a biological sequence, or a
spiking pattern from a population of neurons.

The R\'enyi entropy of order $\beta$ is defined as:
\beq
H(\beta)=\frac{1}{1-\beta}\ln\left[\sum_s p(s)^{\beta}\right].
\eeq
This quantity generalizes Shannon's entropy,
\beq
H_1=-\sum_s p(s)\ln p(s),
\eeq
to which it reduces in the limit $\beta\to 1$.
The R\'enyi entropy is associated with a family of diversity indices, defined as:
\beq
D(\beta)=\exp[H(\beta)]={\left(\sum_s p(s)^\beta\right)}^{\frac{1}{1-\beta}}.
\eeq
This quantity can be interpreted as an effective number of states.
When $\beta= 0$, $D(0)$ is just the raw, total number of possible
types in the system. When $\beta=2$, it is equal to the inverse of
Simpson's index, also interpreted as an effective number of types, and
related to the Gini-Simpson index (defined as $1-1/D(2)$), commonly used to measure inequalities.
 When $\beta=1$, $D(1)$ is the exponential of Shannon's entropy, and
 is sometimes called the true diversity. In Shannon's original work \cite{Shannon}, $D(1)$ is the effective number of codewords needed to compress $s$.

We first derive an equivalence between the R\'enyi entropy and the free
energy of statistical mechanics, as already reported in \cite{Baez2011}.
We formally rewrite the probability distribution $p(s)$ as a Boltzmann distribution:
\beq\label{boltzman}
p(s)=\frac{1}{Z_1}e^{-E(s)},
\eeq
where the temperature is set to $k_BT=1$ by definition. For example,
this mapping can be realized by defining $E(s)\equiv -\ln P(s)$ and
$Z_1=1$, but to keep things general we will assume an arbitrary
$Z_1$. We define the free energy at unit temperature as $F_1=-\ln
Z_1$. The R\'enyi entropy can be rewritten as:
\beq\label{Frenyi}
H(\beta)=\frac{1}{1-\beta}\ln\left[\sum_s e^{-\beta E(s)+\beta F_1}\right]=\frac{\beta(F_1-F(\beta))}{1-\beta},
\eeq
In this formula, $F(\beta)$ is the usual free energy at inverse temperature $\beta$:
\beq\label{freeenergy}
F(\beta)\equiv -\frac{1}{\beta}\ln Z(\beta),
\eeq
where $Z(\beta)$, the normalization factor of the Boltzmann distribution $p_{\beta}(s)=\exp(-\beta E)/Z(\beta)$ at inverse temperature $\beta$,
\beq\label{Z}
Z(\beta)\equiv \left[\sum_s e^{-\beta E(s)}\right],
\eeq
is called the partition function.

Thus the R\'enyi entropy is closely related to the free energy, after
mapping to the Boltzmann distribution. Note that this mapping is a
definition, and does not follow from physical considerations.

\section*{Abundance distribution and micro-canonical entropy}

There also exists a rigorous analogy between the density of states and the abundance distribution \cite{Mora2011a}. The abundance distribution is defined as the distribution over $p$ itself:
\beq
\rho(p)=\sum_s \delta[p(s)-p],
\eeq
where $\delta(\cdot)$ is Dirac's delta function. It is more convenient
to work with the cumulative density of $p$, as it is not plagued with
Dirac deltas, and is invariant under reparametrization:
\beq
C_p(p)=\sum_s \Theta[p(s)-p],
\eeq
where $\Theta(x)$ is the Heaviside function, equal to $1$ for $x\geq
0$ and $0$ otherwise.

The cumulative distribution of abundances is
related to another representation of diversity, the rank-frequency
curve or Zipf's plot \cite{Zipf}. In this representation, the system's states are ranked from most abundant to least abundant, and their abundance shown as a descreasing function of the rank.
The rank of a given abundance $p$ is given exactly by
$C_p(p)$. Hence, rank-frequency graphs are simply plots of $p$ versus
$C_p(p)$. In other words, they represent the inverse function of the cumulative abundance distribution.

Since $p$ and $E$ are related by the Boltzmann distribution \eqref{boltzman}, we can equivalently define the cumulative density of states, which counts all states under a given temperature $E$.
\beq
C_E(E)=\sum_s \Theta[E-E(s)].
\eeq
This cumulative distribution is related to the cumulative distribution of $p$ through $C_E(E)=C_p(e^{-E}/Z_1)$.
The usual density of states,
\beq
\rho(E)=\sum_{s} \delta(E-E(s)),
\eeq
 is obtained as $dC_E(E)/dE$. In order to avoid issues of definition
 with Dirac delta functions, we define a cumulative micro-canonical
 entropy as $S(E)=\ln C_E(E)$, rather than the usual micro-canonical
 entropy. In this definition, for ease of notation we implictly take
 the Boltzmann constant to
 be $k_{B}=1$.

\section*{Equivalence of ensembles}

{Following textbook statistical mechanics,} the partition function \eqref{Z} can be rewritten entirely as a function of the density of states:
\beq\label{laplace}
\begin{split}
Z(\beta)=&\int dE \rho(E) e^{-\beta E} = \beta \int dE C_E(E) e^{-\beta E}\\
&=\beta\int dE e^{S(E)-\beta E},
\end{split}
\eeq
where we have used integration by parts in the second equality. In
other words, $Z(\beta)$ is the Laplace transform of the density of states. 

In the standard thermodynamic limit, where both the entropy and energy
are assumed to be extensive, $S(E) \sim E \sim N$, where $N$ is the system's size, this integral can be approximated by its saddle point, an approximation also known as Laplace's method:
\beq
Z(\beta) \approx \beta{\left(\frac{\pi}{|S''(E^*)|}\right)}^{1/2}e^{S(E^*)-\beta E^*},
\eeq
where $E^*$, which maximizes the term in the exponential in Eq.~\ref{laplace}, is given by the standard thermodynamic relation:
\beq\label{dSdE}
\left.\frac{dS}{dE}\right|_{E*}=\beta,
\eeq
or more classically $dS/dE=1/T$.
The free energy (Eq.~\ref{freeenergy}) then reads:
\beq\label{free}
F(\beta)=E^*-\frac{1}{\beta} S(E^*) -\frac{\ln(\beta)}\beta -\frac{\ln(\pi)}{2\beta}+\frac{\ln(|S''(E^*)|}{2\beta}.
\eeq
In the thermodynamic limit the last three terms are subextensive (scaling sublinearly with the system's size) and
therefore dropped, reducing to the usual definition of the
free energy, $F=E-TS$.
Then the Massieu potential (also called the Helmholtz free entropy)
$\Phi(\beta)=-\beta F(\beta)$ and the micro-canonical entropy $S(E)$ are related by a Legendre transform:
\bea
\Phi(\beta)&=&\mathrm{extr}_E[S(E)-\beta E],\label{legendre}\\
S(E)&=&\mathrm{extr}_\beta[\Phi(\beta)-\beta E],
\eea
in which $E$ and $\beta$ are conjugate variables. These relations define
the equivalence between the micro-canonical and canonical ensembles,
which is valid as long as $S(E)$ is a concave
function \cite{Touchette2004}. In this equivalence, different inverse temperatures
$\beta$ are used to sample states of different typical
energies, acting as a large-deviation parameter. {These
  relations formally follow from the Boltzmann distribution in the
  thermodynamic limit, and are the same as in standard thermodynamics.}

The saddle-point approximation is more than a computational trick. It
also impies that, in the thermodynamic limit, the measure is dominated
by just a few states that all have pratically the same energy
$E^*$. There are of the order of $\exp[S(E^*)]$ such states, which
each have roughly the same probability $\exp(-\beta
E^*)/{Z(\beta)}=\exp[-S(E^*)]$. Their entropy is then given by
Boltzmann's formula:
\beq\label{entropy}
H_1[p_\beta]=-\sum_s p_\beta(s)\ln p_\beta(s)\approx \ln[e^{S(E*)}]= S(E^*),
\eeq
where $H_1[p_\beta]$ is the canonical entropy at inverse temperature $\beta$, not to be confused with the R\'enyi entropy $H(\beta)$.
The result of Eq.~\ref{entropy} can be shown more rigorously by using the exact identity:
\beq
H_1[p_\beta]=\beta[\<E\>_\beta-F(\beta)],
\eeq
with $\<x\>_\beta=\sum_s p_\beta(s)x(s)$, and by showing $\<E\>_\beta\approx E^*$ using Laplace's method.

\section*{Legendre construction}

\begin{figure}
\noindent\includegraphics[width=1\linewidth]{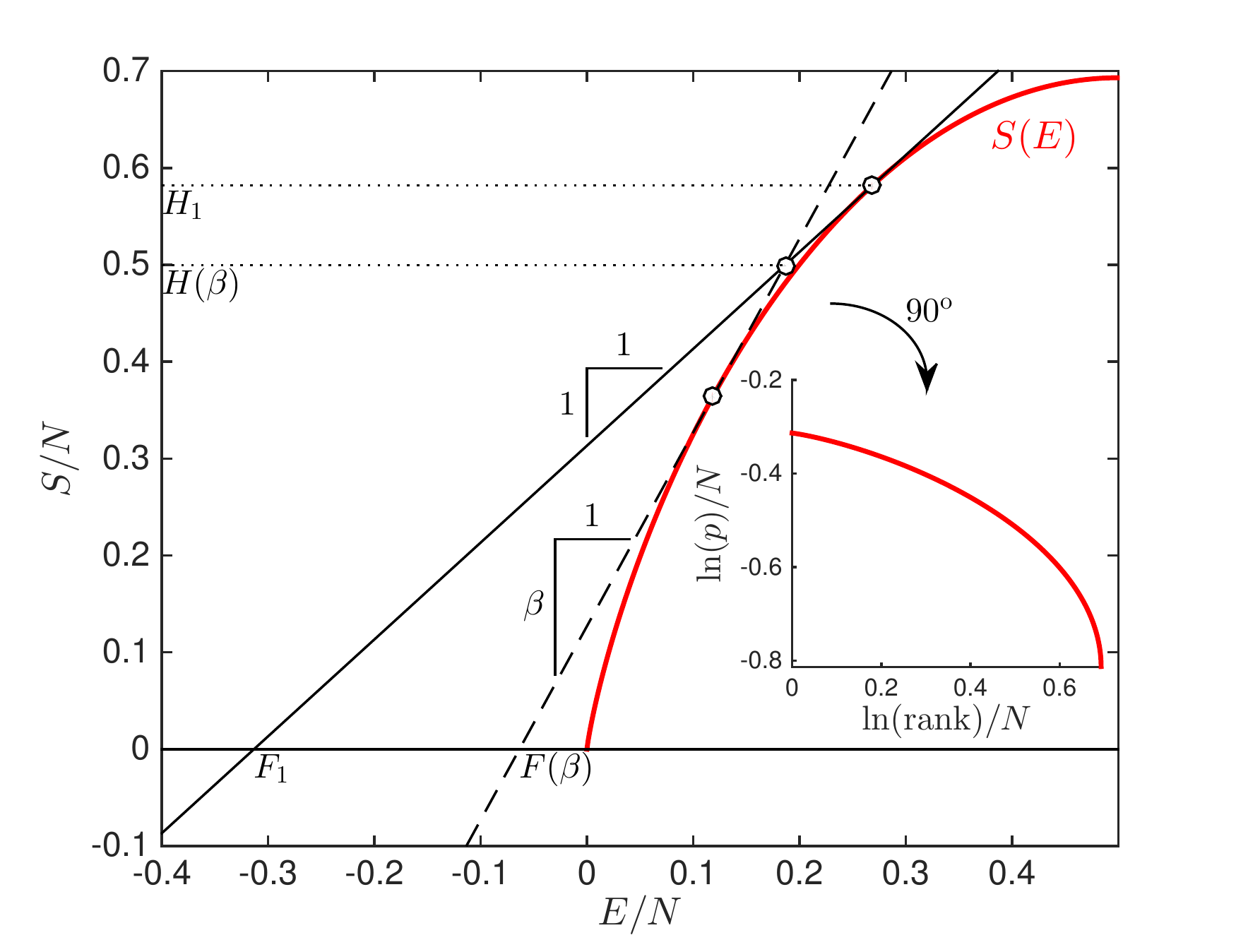}
\caption{Geometric construction of the R\'enyi entropy from the
  density of states. In the classical Legendre construction, the free energy $F(\beta)$ is obtained as the intersection
  of the tangent to the micro-canonical entropy curve $S(E)$ (in red) of slope
  $\beta$, where $\beta$ is the inverse
  temperature, and the abscissa. The R\'enyi entropy of order $\beta$,
  $H(\beta)$, is obtained as the  intersection between the tangents of
  slope $1$ and $\beta$, projected onto the ordinate. Inset: the
  micro-canonical entropy curve is equivalent, up to a 90${}^{\rm o}$ rotation, to the rank-frequency
  curve represented on a logarithmic scale.
}
\label{cartoon}
\end{figure}

The Legendre transform \eqref{legendre} can be constructed
geometrically, as illustrated by Fig.~\ref{cartoon}. In this
construction, $F(\beta)$ is obtained as the intercept of the tangent
to $S(E)$ of slope $\beta$ {(dashed line in Fig.~\ref{cartoon})} with the abscissa. To see this, we write
the condition for $E^*$ at the point where the tangent of slope
$\beta$ touches the $S(E)$ curve, $dS/dE=\beta$, which is exactly Eq.~\ref{dSdE}. The equation defining the tangent is then given in $(E,S)$ space by:
\beq
S=S(E^*)+\beta(E-E^*).
\eeq
Solving in $E$ for the intercept with the abscissa,
$S=0$, gives $E^*-S(E^*)/\beta=F(\beta)$, which is the result of
Eq.~\ref{free} up to the sub-extensive terms.

We can generalize this construction to the R\'enyi entropy, which is
obtained as the intersection of two tangents to $S(E)$, of slopes 1
and $\beta$ respectively. To verify this assertion, one writes the
system of two linear equations parametrizing these two tangents in the
$(S,E)$ space:
\bea
S&=&E-F_1,\\
S&=&\beta[E-F(\beta)].
\eea
The solution to these two equations in $S$ is $\beta(F_1-F(\beta)/(1-\beta)$, which is exactly the R\'enyi entropy $H(\beta)$ according to \eqref{Frenyi}.

As already mentioned, the R\'enyi entropy reduces to the classical Shannon or Gibbs entropy, $H_1$, for $\beta= 1$:
\beq
\lim_{\beta\to 1} H(\beta)= \left.\frac{dF}{d\beta}\right|_{\beta=1}=H_1.
\eeq
This limit can also be undertood geometrically. When $\beta\to 1$, the
intersection between the two tangents tends to the point of tangency
of slope 1,
$dS/dE|_{E^*}=1$, where $S(E^*)$ is equal to the Shannon entropy $H_1$ (Eq.~\ref{entropy}).

\section*{From the abundance distribution to R\'enyi entropy: a geometric approach}
{Now that we have derived analogous
  relations to standard thermodynamics,
we can use the geometric representation of
  the Legendre transform to read off diversity measures from data.}
The Legendre construction can be transposed into the language of the
abundance distribution, provided that this distribution is
appropriately represented as a rank-frequency curve. Recall that
$S(E)=\ln C_E(E)$, where $C_E$ is the rank of states of energy $E$,
ordered from the lowest to the highest energy, {\em i.e.} from the most
frequent to the least frequent state. On the other hand, $E=-\ln p+F_1$, where $p$ is the frequency. Thus, the micro-canonical entropy function, $S$ {\em vs.} $E$, and the rank-frequency relation in logarithmic scale, $\ln(p) $ {\em vs.} $\ln({\rm rank})$, are exactly equivalent up to a $90^{\rm o}$ rotation, as illustrated in the inset of Fig.~\ref{cartoon}.

Thanks to this equivalence, the Legendre construction described above can be applied
directly to the rank-frequency curve plotted on a log-log scale.
We
illustrate such a consctruction with the distribution
of generation probabilities of T-cell receptor beta chains
\cite{Murugan2012}. DNA sequences $s$ coding for the beta chain of
T-cell receptors, which are involved in recognizing pathogens, are
generated according to the probability distribution $p(s)$, which was
inferred from the data. For each generated sequence, its probability {of generation}
$p$, and thus its energy $E=-\ln p$, is also output by the model. This
allows us to compute empirically the probability distribution of $E$ under the model, which is proportional to the
number of states with a certain energy multiplied by their probability $p=e^{-E}$,
$P(E)\propto\rho(E) e^{-E}$, from which $\rho(E)$ and then $C_E(E)=\int_0^E
dE'\rho(E')$ are obtained. Note that this distribution is
synthetically created from the model of generation by drawing random, independent
sequences. It is distinct from clone-size distributions usually found
in the literature \cite{Mora2010,Zarnitsyna2013,Desponds2016}, which
have a clonal structure and are not
made of independent samples. Also note that this ensemble has no natural
thermodynamic limit, because sequences have a finite length. It
makes for a good test case for our method in a real-world example.

\begin{figure}
\noindent\includegraphics[width=1\linewidth]{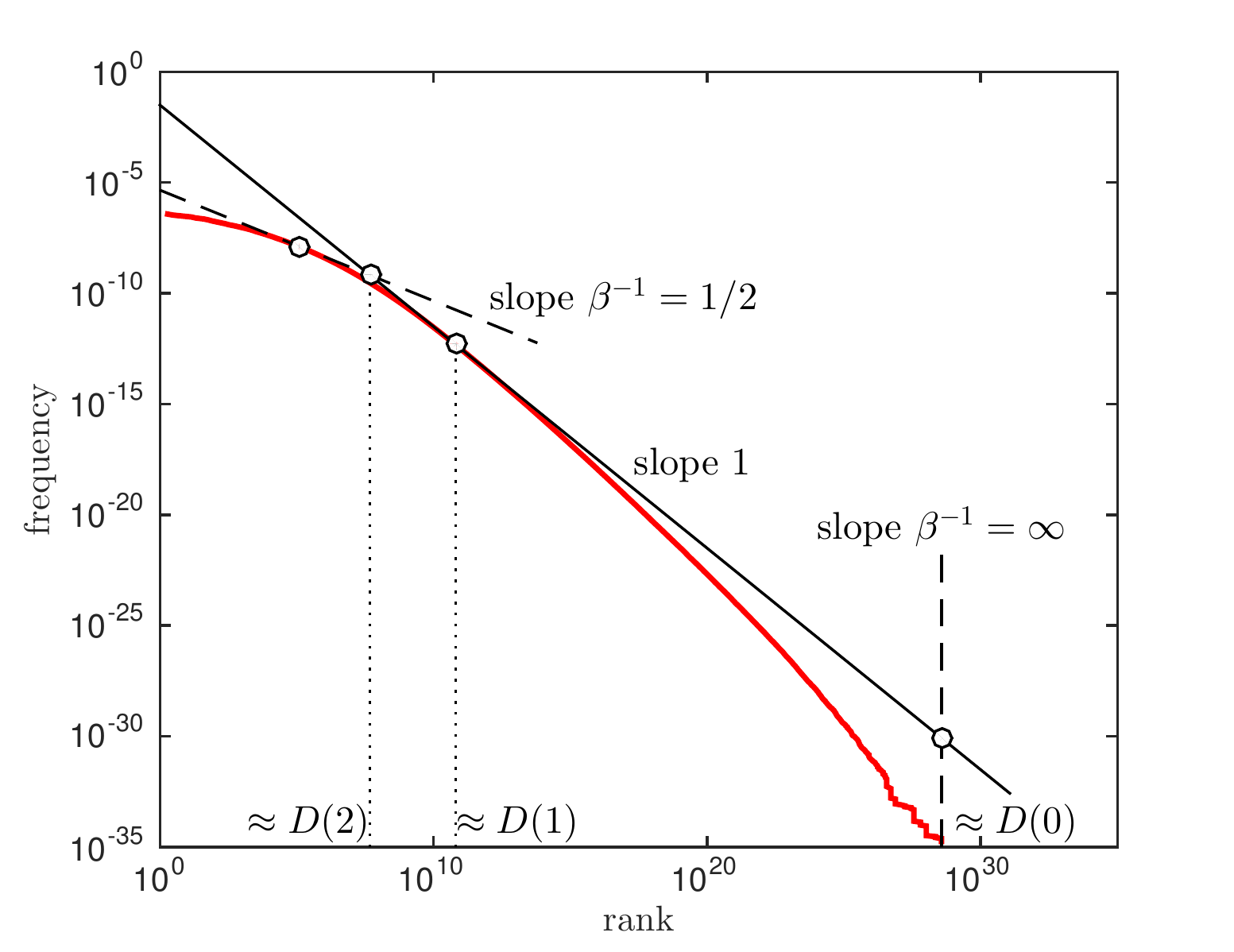}
\caption{Illustration of the Legendre construction of the R\'enyi
  entropy on the rank-frequency curve of randomly generated T-cell
  receptor beta chains \cite{Murugan2012}. The construction is identical to that of
  Fig.~\ref{cartoon}, with slope $\beta$ replaced by slope
  $-\beta^{-1}$ because of the rotation. The projection onto the rank
axis gives an approximation to the diversity index of order $\beta$, $D(\beta)=\exp[H(\beta)]$.}
\label{construction}
\end{figure}

The rank-frequency plot is represented in Fig.~\ref{construction} in a
double logarithmic scale. Following the previous arguments, in this representation the diversity index $D(\beta)=e^{H(\beta)}$
can be approximated by:
\begin{enumerate}
\item drawing the tangent of slope -1 
  (black solid line) to the rank-frequency curve;
\item drawing the tangent of slope $-\beta^{-1}$ (dashed line) to the same
  curve;
\item projecting the intersection point between these two lines onto
  the rank axis.
\end{enumerate}

The tangency point of the tangent of slope 1 gives the true
diversity index $D(1)=\prod_s p(s)^{-p(s)}$, {\em i.e.} the exponential of
Shannon's entropy. In Fig.~\ref{construction} we illustrate the example of
$\beta=2$, which {allows us to read of} 
Simpson's inverse index $D(2)=1/\sum_s
p(s)^2$. The exact values for these two quantities, $D(1)=4.9\cdot 10^{13}$ and
$D(2)=3.4\cdot 10^9$, are roughly approximated, although underestimated, by
the construction.

In the true thermodynamic limit, which is not strictly realized but approached in this
example, the diversity measure $D(\beta)$ is effectively dominated by just a fraction of
sequences whose rank is close to $D(\beta)$ (on a logarithmic scale), according to Laplace's
approximation. A consequence of this concentration is that different diversity
measures, such as Shannon's entropy or Simpson's index, may in fact be 
determined by entirely distinct sequences.

The construction allows for a quick assessment of whether the sampling depth can
support the estimation of the R\'enyi entropy $H(\beta)$, and its associated
diversity $D(\beta)$, for a given index $\beta$. When the tangent of
slope $-\beta^{-1}$ touches the curve towards its end, where states are
becoming rare and may have been observed only once, it is probably safe to assume that the R\'enyi entropy
cannot be reliably computed from the data, because it is determined by
states which have not been sampled well. This limitation applies to the Legendre
construction as well as to any other estimate of Renyi's entropy.
Such a diagnosis indicates
which diversity measure might be appropriate to use in a given
context, depending on the shape of the rank-frequency curve.

An extreme case is
$\beta=0$: the tangent of slope $-\beta{\color{red}^{-1}}=-\infty$ intersects with
the rank-frequency curve at the maximal possible rank, which is also
the total number of sampled types. In most cases (as in this one) this
maximal rank does not represent well the true total diversity, $D(0)$, which should also
include unseen types. A similar underestimation is expected to happen
for finite values of $\beta$ for which the tangent is ill-defined.

\section*{Singular cases}

It is interesting to consider what happens to the R\'enyi entropy when
the rank-frequency relation is locally a power law. In the
micro-canonical framework, a power law in the cumulative density of
abundances \cite{Mora2011a},
\beq
C(p)\propto \frac{1}{p^\alpha},
\eeq
translates into a linear
density of states, $S(E)=S_0+\alpha (E-E_0)$. This behaviour, as long
as it spans an extensive range of energies, leads to a discontinuity in the derivative of
$F(\beta)$ at $\beta=\alpha$.
For $\alpha\neq 1$, Eq.~\ref{Frenyi} implies that the derivative of $H(\beta)$ 
exhibits the same kind of discontinuity at $\beta=\alpha$.

For $\alpha=1$ the
discontinuity is of a different nature. Eq.~\ref{Frenyi} can be expanded around
$\beta=1$ as:
\beq
H(\beta)\approx H_1+\frac{\beta-1}{2}\left.\frac{d^2F}{d\beta^2}\right|_{\beta=1},
\eeq
hence:
\beq
\left.\frac{dH}{d\beta}\right|_{\beta=1}=\frac{1}{2}\left.\frac{d^2F}{d\beta^2}\right|_{\beta=1}.
\eeq
Therefore, the discontinuity in $dF(\beta)/d\beta$ at $\beta=1$
translates into a discontinuity in $H(\beta)$ itself. Again, this
discontinuity can be seen geometrically. Let us assume that $S(E)=S_0+(E-S_0)$ over a
range $(E_1,E_2)$. The tangent of slope 1 coincides with $S(E)$
throughout this range.
As a result, the intersection between the tangent
of slope $\beta$ jumps from $E_1$ to $E_2$ as $\beta$
crosses $1$, causing $H(\beta)$ to jump from $S(E_1)$ to $S(E_2)$.

This kind of singularity not only implies discontinuities in the R\'enyi entropy or its derivatives, but also suggest that the entropy may ill-defined or hard to estimate when $\alpha=1$. In that case, a whole range of $(S,E)$ pairs, instead of a single point, are candidates for the tangency point between the line of slope 1 and the micro-canonical entropy $S(E)$. The entropy is ultimately determined by corrections that are ignored in the thermodynamic limit.

The micro-canonical entropy need not be strictly linear over a portion
of energies for a discontuinuity to occur. In fact, any convexity in
$S(E)$ is predicted to produce the same effect \cite{Touchette2004}.

For this reason, caution should be used when dealing with
distributions that look like a power law, or are not concave in
logarithmic scale. Not only may the Legendre
construction be unreliable, but so may other more direct estimates
of the R\'enyi entropy, {since the system lacks a characteristic
  energy scale}. Interestingly, several abundance distributions in biology
have been reported to follow power laws with exponent $\alpha=1$
\cite{Mora2011a,Stephens2013,Tkacik2015}, for which the R\'enyi
entropy is expected to have a discontinuity.

\section*{Discussion}
In this paper we have made an explicit link between
classical representations of diversity in ecology
and other fields, and the
framework of classical statistical mechanics. This mapping allows one
to bring many quantities coming under many different names
-- species abundance distribution, clone-size
distribution, frequency spectrum, Shannon entropy, R\'enyi entropy,
Simpson's index, {\em etc.} -- within a common
framework. It provides an quick an easy way to {simply
  read off} 
 diversity directly
from rank-frequency plots.

Our geometric construction assumes the thermodynamic limit, which may
not be satisfied or even well defined. For instance, the
distribution of abundances prediced by a neutral model, or Fisher's log-series
$P(n)\propto\alpha^n/n$ \cite{Fisher1943}, does not admit a natural definition of system's size, and thus has no well defined
thermodynamic limit. The same goes for Pareto distributions $P(n)\propto
n^{-a}$. In these cases where no thermodynamic limit exists, the
Legendre construction is no substitute for a direct estimate, but may
still give a reasonable guess. It can also hint whether such a direct
estimate is possible at all, by identifying the range of frequencies
or abundances that are expected to dominate the diversity measure.

Depending on how well sampled the distribution is, different
orders $\beta$ of the R\'enyi entropy may be appropriate.
The proposed framework can aid in choosing the right
measure depending on the data. In general, the less well sampled the data is, the
higher the order should be chosen.
For instance, Simpson's index is less
sensitive to poor sampling than the Shannon entropy, which itself is
easier to estimate from the data than the total number of
states. Ultimately, the particular form of the abundance distribution
should be examined to decide which measure can or should be used.

This work was supported in part by grant ERCStG n. 306312, and by the
National Science Foundation under Grant No. NSF PHY11-25915 through
the KITP where part of this research was done.

\bibliographystyle{pnas}

\end{document}